\documentclass{aastex62}

\usepackage{mathtools}
\usepackage{epsfig}
\usepackage{comment}
\setcitestyle{citesep={,}}

\usepackage{ulem}

\renewcommand{\thefootnote}{\fnsymbol{footnote}}

\received{March 16,  2019}
\revised{August 14, 2019}
\accepted{October 11, 2019}
\submitjournal{ApJ}

\shorttitle{\Fermi-LAT analysis \& hadronic modeling of \J}
\shortauthors{Cui et al.}

\newcommand\simlt{\lower.5ex\hbox{$\; \buildrel < \over \sim \;$}}

\newcommand{\J}{{\rm HESS J1731-347}}
\newcommand{\Fermi}{{\sl Fermi}}

\begin{document}

\title{Is the SNR HESS J1731-347 colliding with molecular clouds?}
\correspondingauthor{Yudong Cui, Ruizhi Yang, P.H. Thomas Tam}
\email{cuiyd@mail.sysu.edu.cn, ryang@mpi-hd.mpg.de, tanbxuan@mail.sysu.edu.cn}

\author{Yudong Cui}
\affil{School of Physics and Astronomy, Sun Yat-Sen University, Guangzhou, 510275, China}

\author{Ruizhi Yang}
\affiliation{Max-Planck-Institut f{\"u}r Kernphysik, P.O. Box 103980, 69029 Heidelberg, Germany}

\author{Xinbo He}
\affiliation{School of Physics and Astronomy, Sun Yat-Sen University, Guangzhou, 510275, China}

\author{P.H. Thomas Tam}
\affiliation{School of Physics and Astronomy, Sun Yat-Sen University, Guangzhou, 510275, China}

\author{Gerd P\"uhlhofer}
\affiliation{Institut f{\"u}r Astronomie und Astrophysik, Eberhard Karls Universit{\"a}t T{\"u}bingen, Sand 1, D 72076 T{\"u}bingen, Germany}


\begin{abstract}
The supernova remnant (SNR) \J~is a young SNR which displays a non-thermal X-ray and TeV shell structure. A molecular cloud at a distance of $\sim$3.2\,kpc is spatially coincident with the western part of the SNR, and it is likely hit by the SNR. The X-ray emission from this part of the shell is much lower than from the rest of the SNR. Moreover, a compact GeV emission region coincident with the cloud has been detected with a soft spectrum. These observations seem to imply a shock-cloud collision scenario at this area, where the stalled shock can no longer accelerate super-TeV electrons or maintain strong magnetic turbulence downstream, while the GeV cosmic rays (CRs) are released through this stalled shock. To test this hypothesis, we have performed a detailed \Fermi-LAT reanalysis of the \J\ region with over $9$ years of data. Two distinct GeV components are found, one displaying a soft spectrum is from the compact GeV emission region, the other one displaying a hard spectrum is from the rest of the SNR (excluding the cloud region). A hadronic model involving a shock-cloud collision scenario is built to explain the $\gamma$-ray emission from this area. It consists of three CR sources: run-away super-TeV CRs that have escaped from the fast shock, leaked GeV CRs from the stalled shock, and the local CR sea. The X-ray and $\gamma$-ray emission of the SNR excluding the shock-cloud interaction region is explained in a one-zone leptonic model. Our shock-cloud collision model explains the GeV-TeV observations from the clouds around \J, i.e.\ a cloud in contact with the SNR and a distant cloud in spatial coincidence to the TeV source HESS J1729-345. We find however that the leaked GeV CRs from the shock-cloud collision do not necessarily dominate the GeV emission from the clouds, due to a comparable contribution from the local CR sea. 
\end{abstract}

\section{Introduction}
\label{Intro}
\J\ was first discovered as an unidentified TeV source by \cite{Ah2008a}. A newly discovered radio SNR (G353.6-0.7) was then found in spatial coincidence to the TeV source \citep{Tian2008}. The following X-ray and TeV observations \citep{Tian2010,Ab2011,Bamba2012} have revealed an X-ray and TeV shell structure of non-thermal radiation from the SNR. An extended TeV structure -- HESS J1729-345 was found Northwest of the SNR. At the center of the SNR, thermal X-ray emission from a central compact object (CCO) has been discovered by \cite{Tian2010,Ab2011}. This CCO is likely a neutron star (NS) \citep{Klochkov2013}, although pulsations have been detected neither in radio nor in X-rays. Upper limits of the GeV emission from the SNR region have been given by \cite{Yang2014,Acero2015b} from a \Fermi-LAT data analysis. Recently, more detailed GeV spectra with power-law indices of -1.7 and -1.8 has been given by \cite{Condon2017} and \cite{Guo2018}, respectively. All of these hard GeV spectra slightly favor a leptonic origin of the $\gamma$-ray emission from the SNR. The lack of thermal X-ray emission of the SNR leads to an upper-limit (90\% confidence level) on the ambient medium density of $\sim0.01$\,cm$^{-3}$ via a shocked medium model \citep{Ab2011}. 

The CO studies together with the X-ray studies by \cite{Ab2011,Doroshenko2017,Maxted2018} have suggested a giant MC of the Scutum-Crux arm ($\sim$3.2\,kpc) to be located right in front of the western part of the SNR, and this giant MC could have already been encountered by the SNR \citep{Doroshenko2017,Maxted2018}. This distance of 3.2\,kpc is also supported by the NS modeling work by \cite{Klochkov2013,Klochkov2015}. Hence, we adopt the distance of 3.2\,kpc in our work, and it leads to a SNR radius of 15\,pc. The age of the SNR is estimated to be 2kyr - 6kyr through modeling several CC SNR scenarios \citep{Cui2016}. These ages are consistent with those derived by the modeling work of a cooling NS \citep{Klochkov2015}. 

In our previous modeling work on \J\ \citep{Cui2016}, run-away super-TeV CRs that have escaped from the SNR shock are at present illuminating the MC at HESS J1729-345 (MC-J1729) through hadronic $\gamma$-ray production.  Additionally, \cite{Cui2016} assumed a leptonic explanation for the $\gamma$-ray emission from the entire SNR region. 
The densest cloud of this entire MC complex at 3.2\,kpc, i.e.\ MC-core, is spatially overlapping with the western part of the SNR. \cite{Cui2016} had put MC-core 100\,pc away from the SNR, in order to prevent the run-away CRs from reaching MC-core, and allow the leptonic model to dominate the $\gamma$-ray emission in the entire SNR image. 

Interestingly, a relatively compact GeV source in spatial coincidence with MC-core was also found in the recent \Fermi-LAT analysis work by \cite{Condon2017}. This source -- named S0 -- shows a power-law index of -2.5, in contrast to -1.7 measured for the entire SNR. However, \cite{Condon2017} did not derive the spectral energy distribution (SED) for S0 in their work. One important open question is whether the emission from S0 (or even from the entire MC-core region) could be of hadronic origin. Further XMM-Newton studies on the SNR by \cite{Doroshenko2017} revealed that the average X-ray surface brightness -- with energies up to 10\,keV -- from the shell in MC-core region is significantly lower than the surface brightness from other regions of the SNR. All of these findings support the hypothesis that this young SNR has already collided in the West with MC-core. Hence, this part of the shock is stalled by MC-core such that it can no longer accelerate super-TeV electrons, nor can it maintain a strong magnetic turbulence downstream. Without the confinement by strong magnetic turbulence at the shock, the low energy CRs are released into nearby MCs through this stalled shock. This hypothesis naturally implies that the GeV-TeV $\gamma$-ray emission of \J\ can be separated into two components, one hadronic from the collision region and one leptonic from the rest of the SNR. 

The work explores the shock-cloud collision hypothesis in the case of \J. The previous \Fermi-LAT analysis work by \citet{Condon2017} only provided a power-law index of MC-core (S0), therefore our main goal in Section~\ref{data} is to subtract the GeV emission of MC-core and obtain the GeV spectra of both MC-core and the rest of the SNR. In Section~\ref{model}, our hadronic model describes how the young SNR is colliding with MC-core and how it releases its GeV-TeV CRs over time into the nearby environment.

\section{\Fermi-LAT Data Analysis}
\label{data}
\subsection{Data preparation}
We select more than nine years of data (MET 239557417 - MET 545548766) observed by \Fermi-LAT for regions around the shell of the supernova remnant \J~ and used the standard LAT analysis software (v11r5p3) \footnote{http://fermi.gsfc.nasa.gov/ssc}. Only events with energy above 1 GeV are used so that the point spread function (PSF) is sharp enough to disentangle multiple spatial components. The region-of-interest (ROI) is selected to be a $10^ \circ \times 10^ \circ$ square centered on the position of J1731-347.  Observations with Rocking angle larger than $52^{\circ}$ are excluded in this analysis. 

In order to reduce the effect of the Earth albedo background, we exclude  the time intervals when the parts of the ROI were observed at zenith angles $> 90^{\circ}$.  The spectral analysis is performed based on the P8R2 version of post-launch instrument response functions (IRFs).  Both the front and back converted photons are selected.

\subsection{FL8Y results}
\begin{figure*}[htb]
\centering
\vspace{-0.cm}
\includegraphics[width=75mm,angle=0]{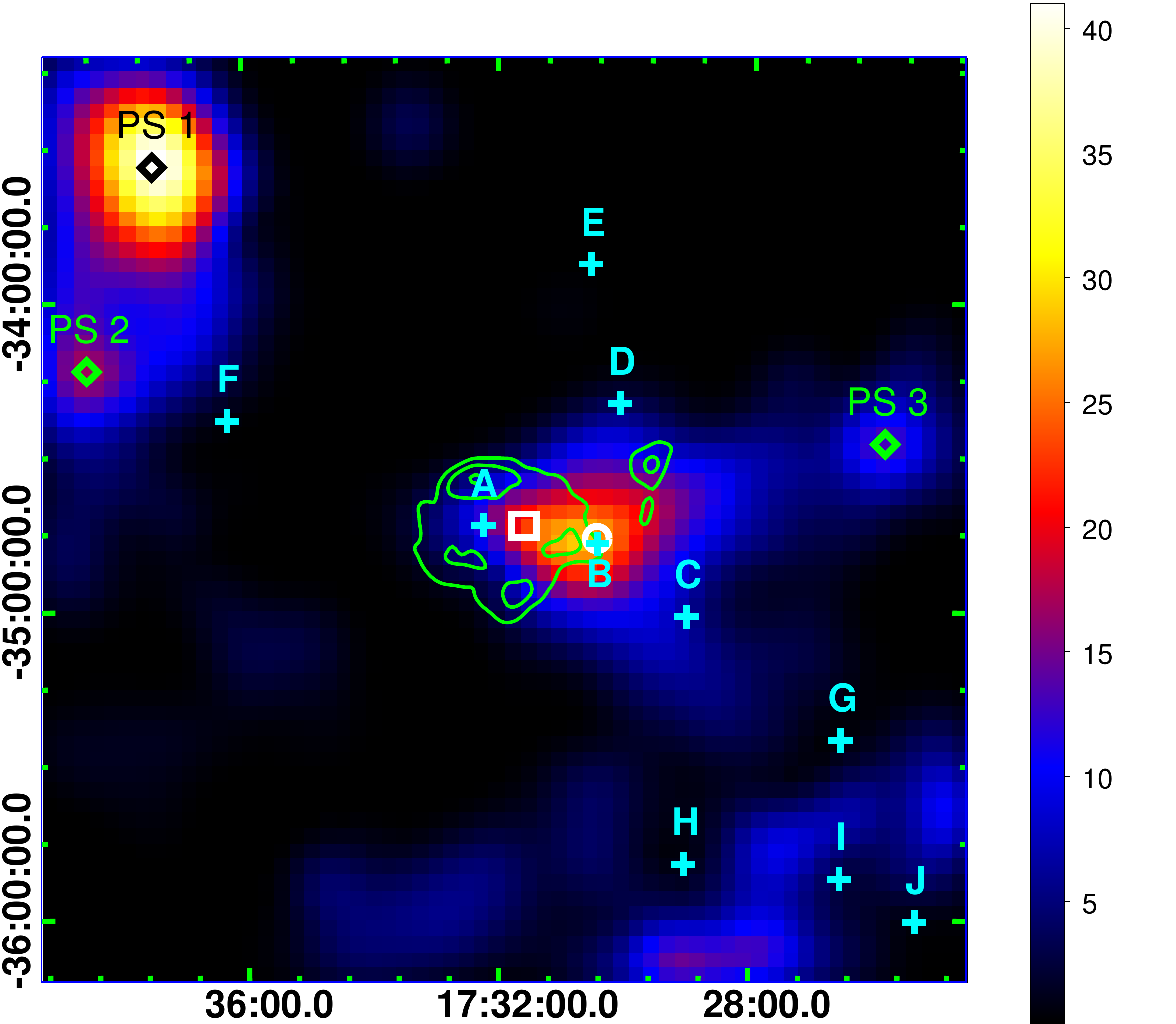} 

\

\hspace{-0.1cm}
\includegraphics[width=78mm,angle=0]{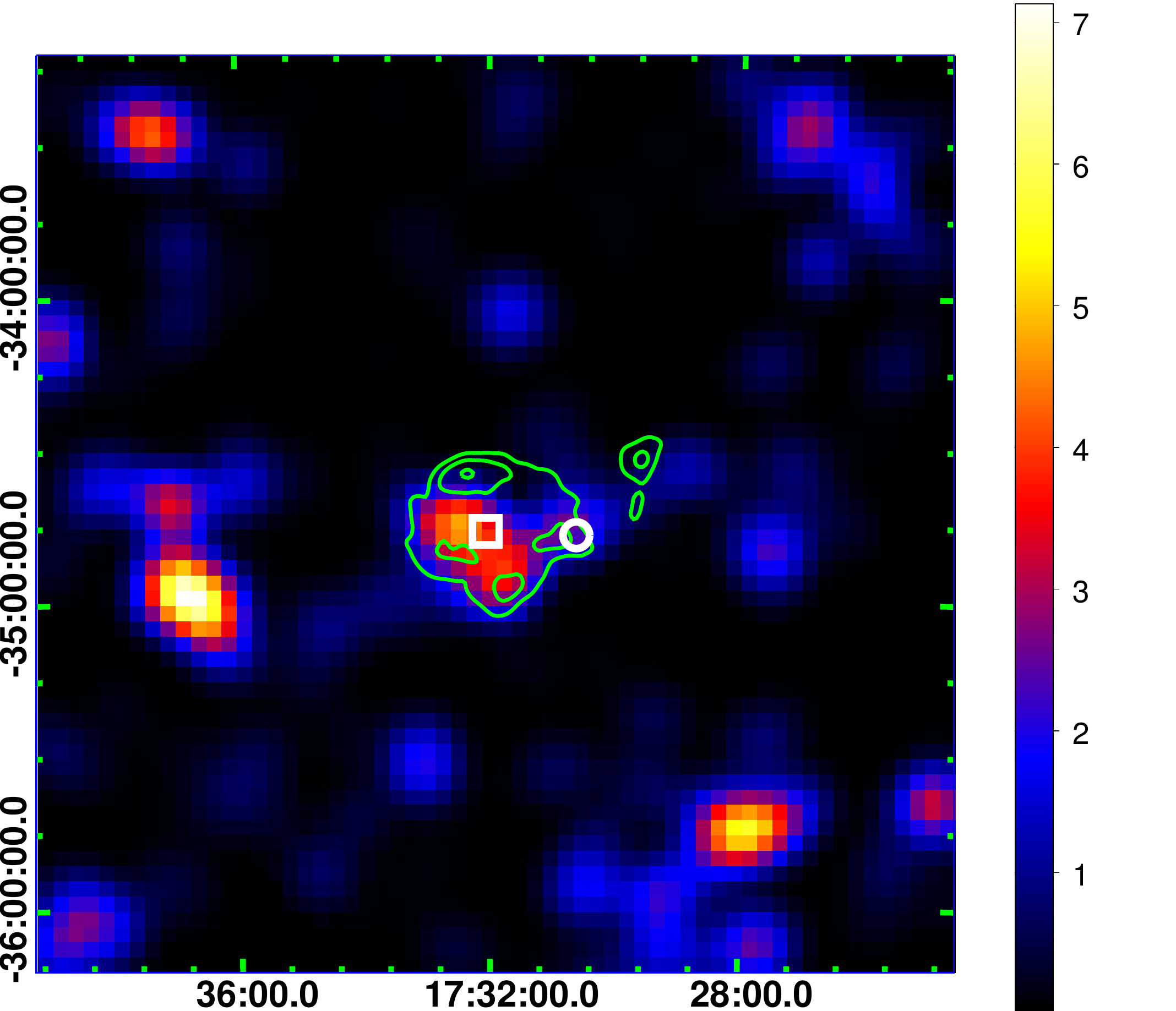}

\

\hspace{-0cm} \includegraphics[width=76mm]{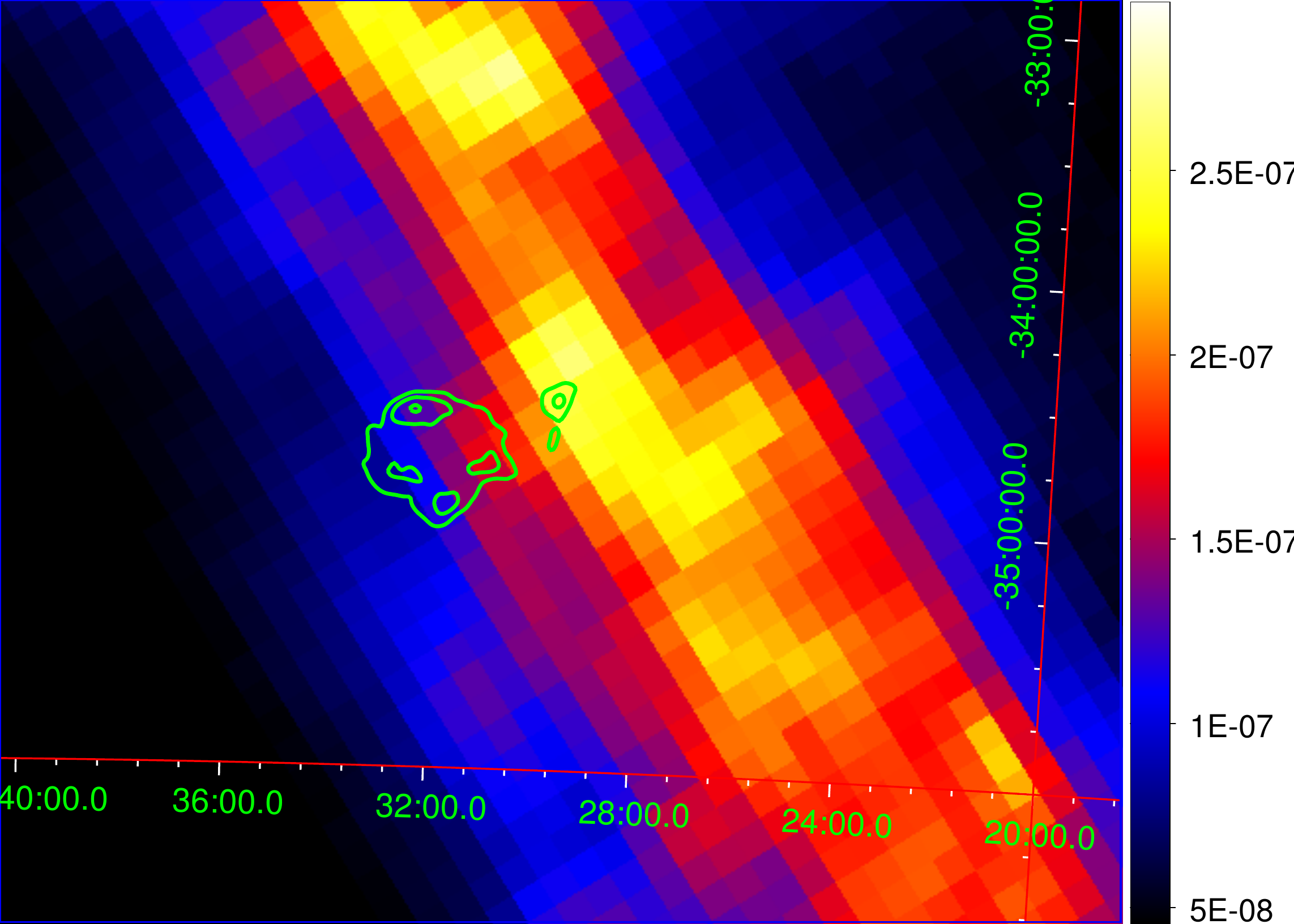}
\caption{Top panel: TS map above 1 GeV.  Middle panel: TS map above 10 GeV. Bottom panel: the counts map of the used diffuse model at 1\,GeV with a unit of $\mathrm{photon\,s^{-1}\,sr^{-1}\,MeV^{-1}\,cm^{-2}}$. In all panels, the x-axis are the Right ascension (RA), and the y-axis are the Declination (Dec); the green contours represent the HESS observations; the three new found point sources, i.e. PS1, PS2, and PS3, are marked as green and black diamonds; the white boxes represent the best fit positions of \J~by using \Fermi-LAT data and the white circles are the positions for the source "S0". The \Fermi~8 years catalog sources (FL8Y) are labeled as cyan crosses, and they are FL8Y\,J1732.2-3443 (A), FL8Y\,J1730.4-3447 (B), FL8Y\,J1729.0-3501 (C), FL8Y\,J1730.0-3419 (D), FL8Y\,J1730.5-3352 (E), FL8Y\,J1736.2-3423 (F), FL8Y\,J1726.5-3525 (G), FL8Y\,J1729.0-3549 (H), FL8Y\,J1726.5-3552 (I), FL8Y\,J1725.3-3600 (J). }
\label{fig:TS}
\end{figure*}

The galactic and isotropic diffuse model provided by the \Fermi~collaboration \footnote{Files: gll\_iem\_v06.fit and iso\_P8R2\_SOURCE\_V6\_v06.txt available at \\ http://fermi.gsfc.nasa.gov/ssc/data/access/lat/BackgroundModels.html} is used in the analysis, and the corresponding counts map of the diffuse model at 1\,GeV is shown in the bottom panel of Fig.~\ref{fig:TS}. \Fermi-LAT 8 years sources are also included, and the parameters for point sources within the ROI are allowed to vary.  We note that \J~ was already included in the 8 years catalog as FL8Y J1732.2-3443. And the source FL8Y J1730.4-3447 are coincide with the source S0 in \cite{Condon2017}.  In Fermi LAT analysis, the likelihood ratio test are used to hypothesis testing. The Test statistics are defined as $TS=-2 ({\rm ln} L_0-{\rm ln} L_1)$, where   $L_0$ and $L_1$ is the likelihood function value in the hypothesis without and with the corresponding source. The source detection significance in $\sigma$ are roughly square root of the TS value \citep{Mattox1996}. The Test statistic maps of the inner $3^{\circ}$ is shown in Fig.~\ref{fig:TS}.   

In deriving the TS maps we do not include the source  FL8Y J1732.2-3443 and FL8Y J1730.4-3447 in the source model.  We also find strong excess to the west and north east of \J, we label them as three new point sources (PS1, PS2, PS3) and add them in the further analysis. For simplicity we fix the position of these additional sources in the maximum of the TS excesses, rather than perform a full likelihood analysis to find the exact position. The influence of the additional sources are checked by simply removing these sources, and those influence are included in the systematic errors of the measured spectra.  We then perform likelihood analysis by varying the position of \J~(FL8Y J1732.2-3443) and find that the best-fit position is at (RA= $262.953^{\circ}$ and Dec=$-34.722^{\circ}$, labeled as white diamond in Fig.~\ref{fig:TS}. The TS value for \J~ is 26 (corresponding to 5.1$\sigma$). We perform the same procedure for S0 (FL8Y J1730.4-3447), the best fit position is (RA=$262.710^{\circ}$ and Dec=$-34.780^{\circ}$, labeled as cyan circle in the upper panel of Fig.~\ref{fig:TS} and the resulted TS value is 27 (5.2$\sigma$). Addition to the $>1$GeV TS map, we have also derived the $>10$GeV TS map, as seen in the middle panel of Fig.~\ref{fig:TS}, in which the soft component -- S0 can be easily told apart from the hard component -- SNR. We also check the extension of \J~by assuming disk templates with different radius and found no improvement of the fit, furthermore, the spectrum of \J~and S0 are significant different, we consider them as independent sources. 

For the spectral analysis  we  applied  \emph{gtlike} in the  energy range [1, 300] GeV and modelled the spectrum of both sources as a power-law function, fixing the position to that found above.   The derived  photon index for \J~ is $1.79 \pm 0.22 (stat) \pm 0.10(sys)$, and  the total flux above 1 GeV is $4.0 \pm 1.9(stat)\pm 0.4(sys) \times 10^{-10} ~\rm cm^{-2} s^{-1}$, while for S0 the index is  $2.42 \pm 0.22 (stat) \pm 0.13(sys)$ and the flux is $1.2 \pm 0.34(stat)\pm 0.14(sys) \times 10^{-9} ~\rm cm^{-2} s^{-1}$.  For the systematic errors we include the errors  coming from the uncertainties of the effective area and the point spread function of LAT \citep{Acero2015a}.  We also vary the normalization of the diffuse background by 6\% to check the influence on the derived  spectra of S0 and \J. Finally we also include uncertainties of the added sources by including or removing the three sources in the likelihood analysis.    

To obtain the spectral energy distribution (SED), we divided the full energy range into 5 logarithmically spaced bands and applied \emph{gtlike} to each of these bands.  The results of this analysis are shown in Fig.~\ref{fig:SED}. All data points have TS values larger than 4, which corresponds to a significance of larger than $\sim2\sigma$.  

\subsection {4FGL results}
\begin{figure*}[htb]
\centering
\includegraphics[height=60mm,angle=0]{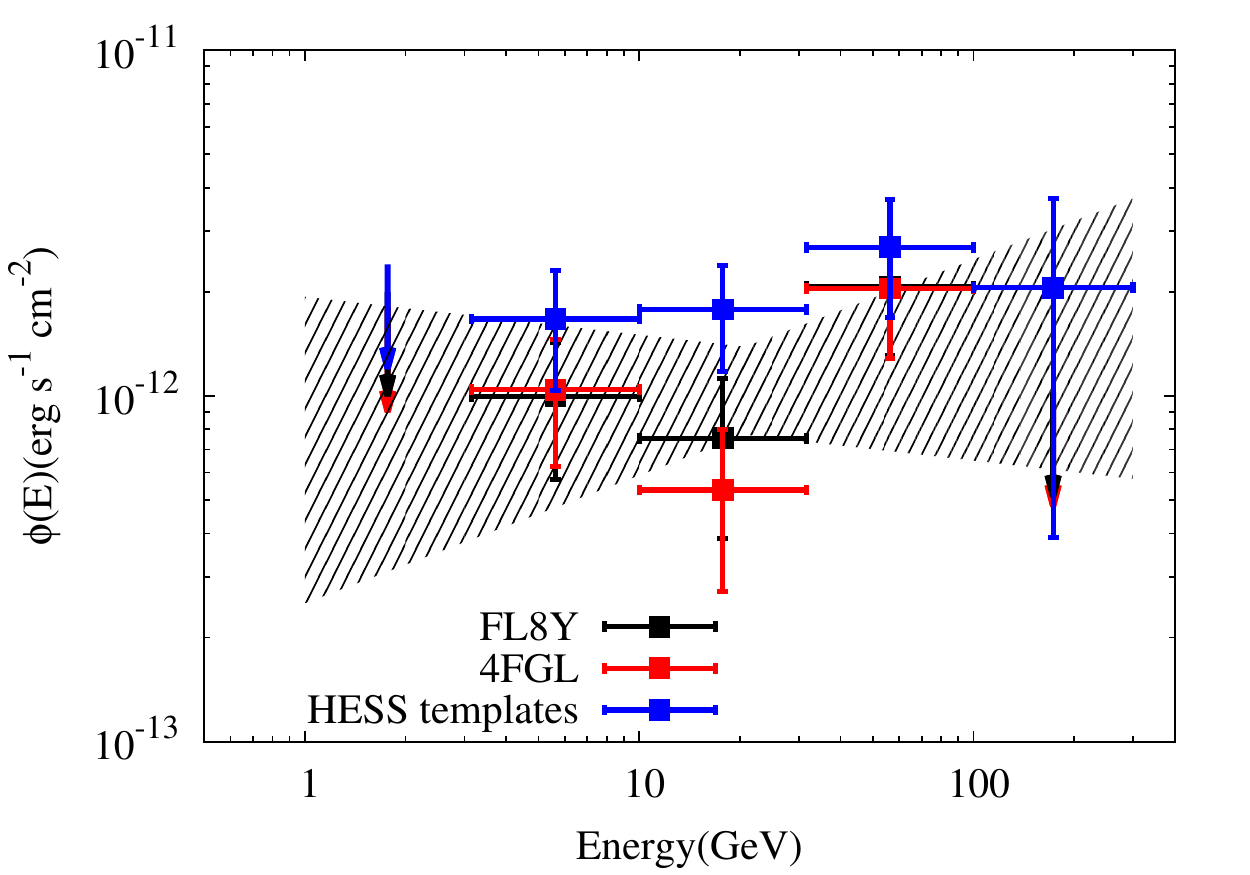}

\includegraphics[height=60mm,angle=0]{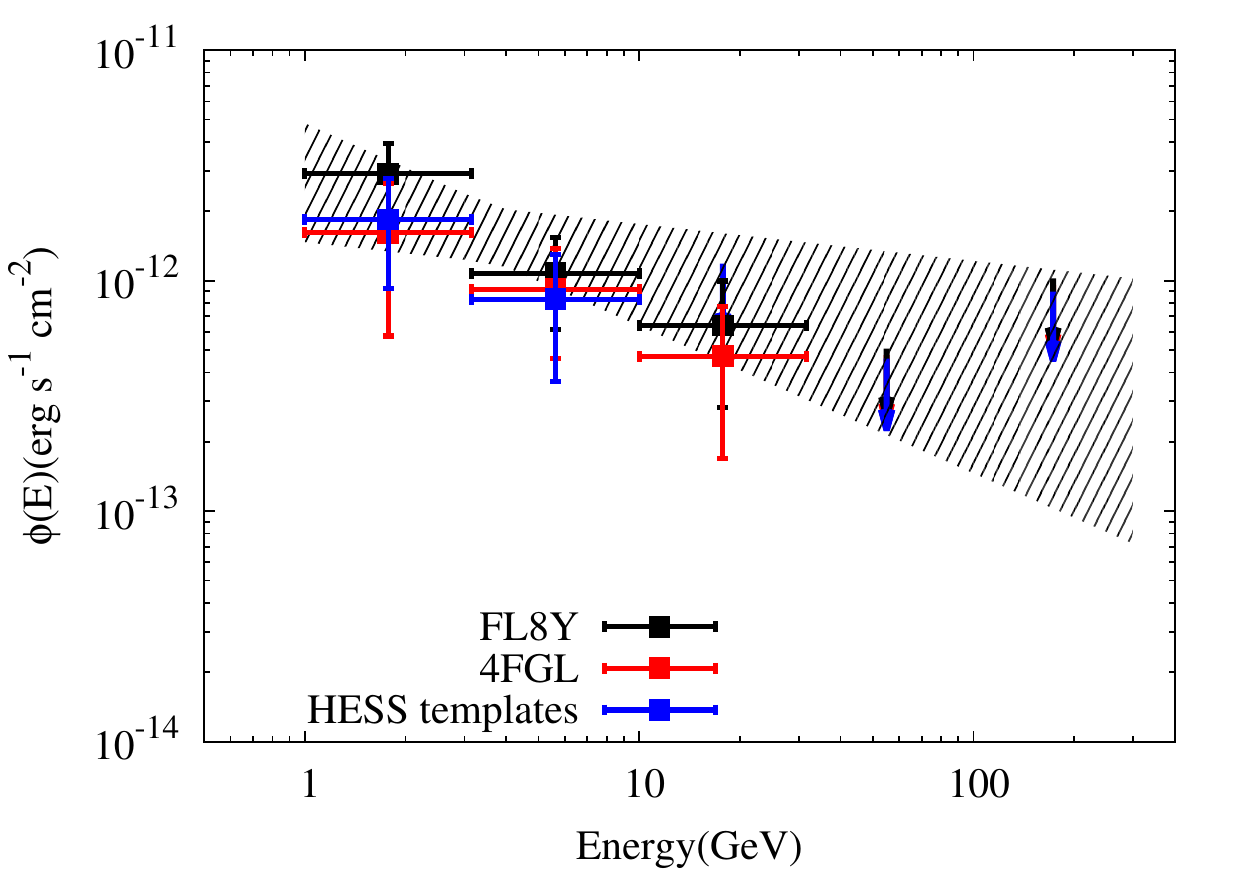}

\

\includegraphics[height=60mm,angle=0]{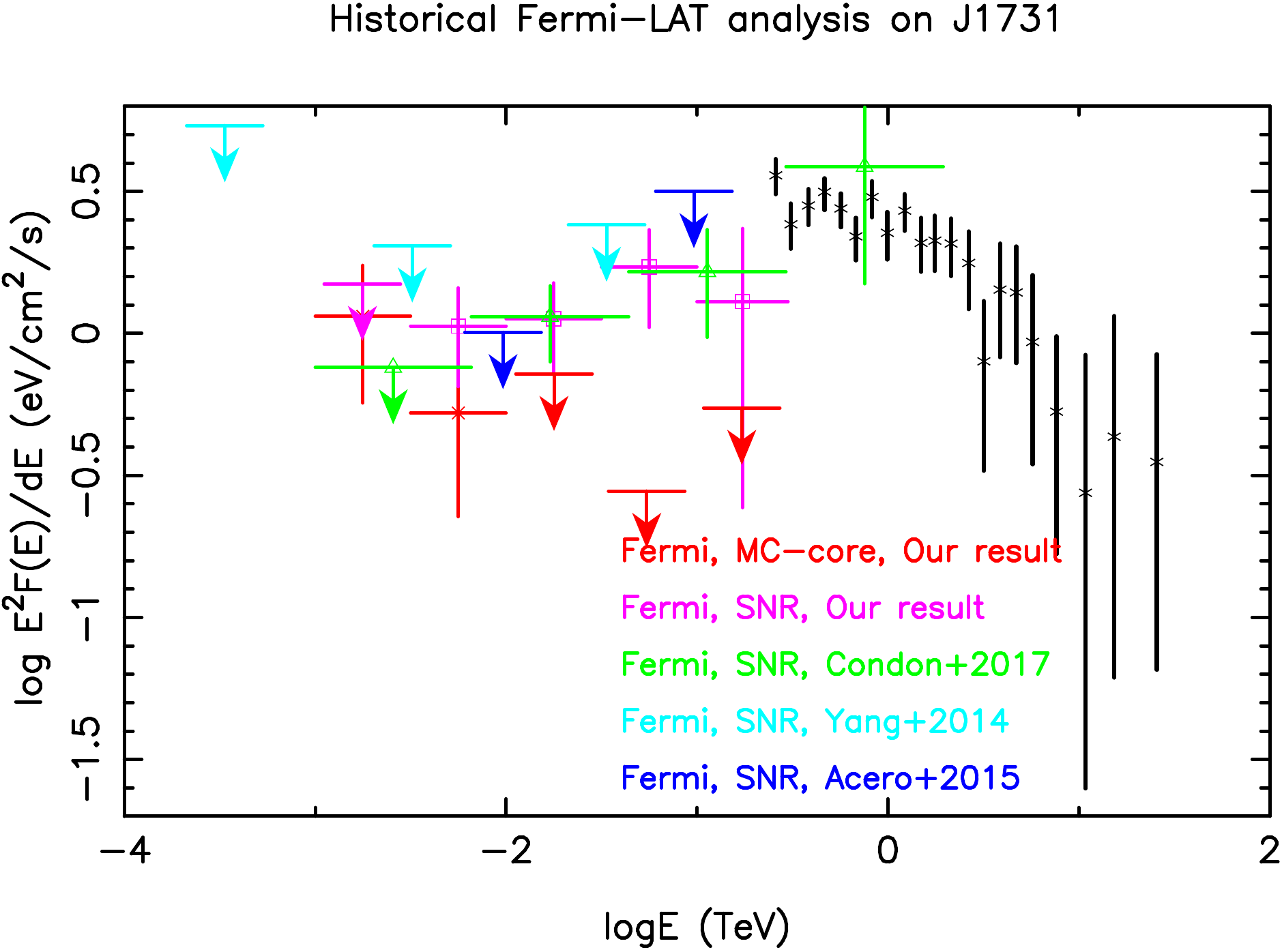}
\caption{The derived GeV spectrum for the SNR \J~(top panel) and S0/MC-core (middle panel). In the top two panels, the spectra using 4FGL catalog are marked in red while the ones using FL8Y are marked in black and grey (for bowtie). The historical \Fermi-LAT analysis of \J~ are shown in the bottom panel, and we only show our results of the HESS template here. The black stars in the bottom panel represents the HESS observation on the SNR \citep{Ab2011}.}
\label{fig:SED}
\end{figure*}

\begin{figure*}[htb]
\centering
\vspace{-0.cm}

\includegraphics[width=90mm,angle=0]{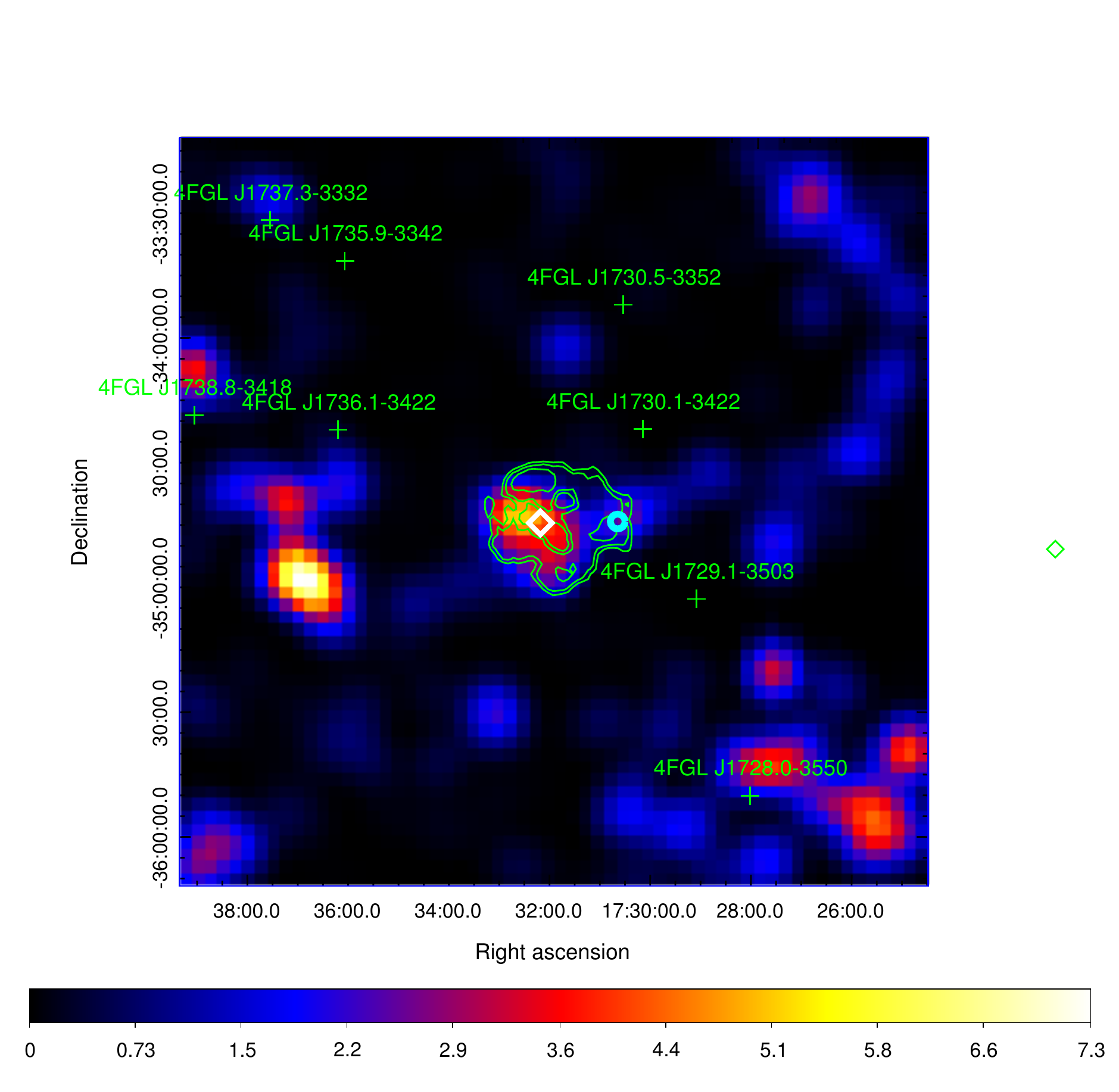}

\caption{ TS map above 10 GeV in the 4FGL analysis.  The green contours represent the HESS observations; the 4FGL  catalog sources are labeled as green crosses; the white diamond labeled the best fit position of \J~ by using \Fermi-LAT data and the cyan circle are the position for the source "S0". }
\label{fig:4fgl}
\end{figure*}

During the process of this work 4FGL catalog and the corresponding diffuse background model {\footnote{Files: gll\_iem\_v07.fit and iso\_P8R3\_SOURCE\_V1\_v01.txt from http://fermi.gsfc.nasa.gov/ssc/data/access/lat/BackgroundModels.html} } have been published. We use the P8R3 data of the same period and the Conda version of Fermi tools to redo the analysis above.  We show the TS map above 10 GeV in this case in Fig.\ref{fig:4fgl}. The point sources corresponding to \J~and S0 are removed from the catalog. Near the positions of our added point source PS1 and PS2, there are three new point sources. Thus for the 4FGL catalog we didn't add any new point source. We perform the similar analysis mentioned above to find the position of \J~and S0. The derived position are  (RA= $262.943^{\circ}$ and Dec=$-34.742^{\circ}$ for \J~and  (RA=$262.636^{\circ}$ and Dec=$-34.815^{\circ}$) for S0, which are slightly different from the ones in FL8Y case but within the error bars. The derived TS value for \J~and S0 are 27 (corresponding to 5.2$\sigma$) and 14 ( 3.5$\sigma$). We also performed the above spectral analysis for the 4FGL case, the results are shown as red data points in Fig.\ref{fig:SED}. Ultimately, the 4FGL results are similar to the FL8Y results, except for S0 which shows a lower flux at low energy band ($<10$\,GeV). This low energy band is dominated by CR sea in our following model, and see relevant discussions in section~\ref{CR_sea}.

\begin{figure*}[htb]
\centering
\vspace{-0.cm}

\includegraphics[width=90mm,angle=0]{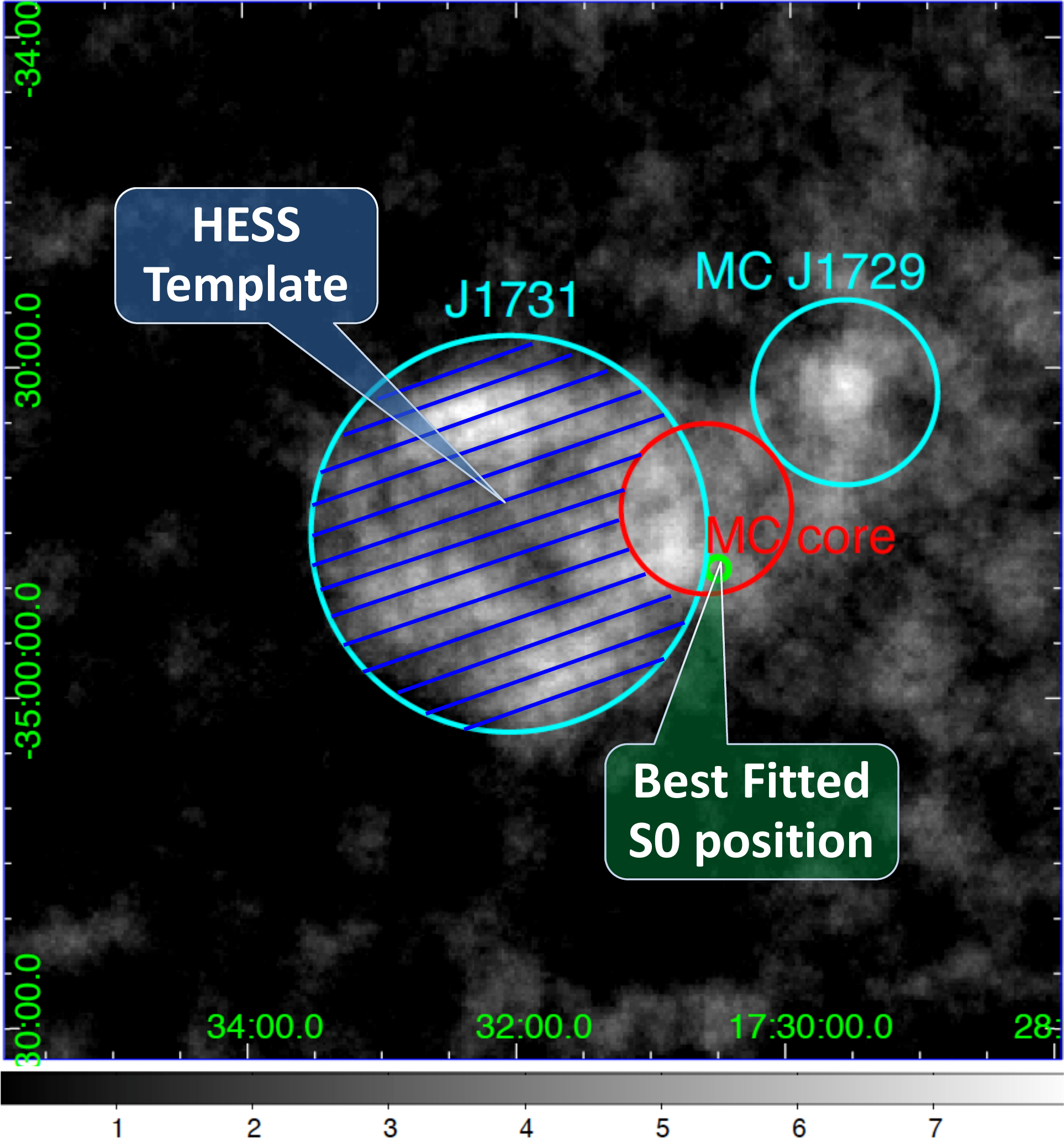}

\caption{ The HESS excess map \citep{Ab2011} as the template for our model-based Fermi-LAT analysis is shown. The HESS template is marked in the shaded area with blue lines. The newly best-fitted position of S0 is marked in green circle. The cyan circles represent the on-regions for \J~\& HESS J1729-345 in the HESS analysis \citep{Ab2011}.}
\label{fig:HESS}
\end{figure*}

To compare the results directly with HESS observations we also use the HESS excess map as the template of \J. Taking into account S0, we subtract a circle (MC-core region) from the HESS excess map (see fig.\ref{fig:HESS}). This analysis template is based on our model (section~\ref{model}), in which the GeV-TeV emission from the HESS template (\J) is leptonic, while the GeV-TeV emission from MC-core (S0) is hadronic. One can also find the definition of MC-core region in our model (section~\ref{MC}). We then perform the likelihood analysis to derive the flux from both \J~and S0. The newly found best-fit position of S0 is RA= $262.629^{\circ}$ and Dec=$-34.808^{\circ}$. In this case the derived TS value for \J~and S0 are 30 (corresponding to 5.5$\sigma$) and 10 (3.1$\sigma$), respectively. The derived SEDs are also shown in Fig.\ref{fig:SED} as blue points. The flux of \J, especially at the $>10$\,GeV energy band, are larger than the point source case due to the larger area of the source, but the total likelihood is not improved.

\subsection{Data analysis summary}

In summary, our results have further supported the recent finding of S0 by \cite{Condon2017}, that the GeV emission from \J~ is likely to be dominated by two components, a hard component at SNR and a soft component at S0. After separating the S0 from the SNR, our GeV flux of the SNR is basically consistent with those historical ones, as seen in the bottom panel of Fig.~\ref{fig:SED}. In the following modeling work, we only show the analysis results of the HESS template for a better model-consistency.

\section{The Hadronic Model}
\label{model}
 Our hadronic model is made of three sub-models, the SNR evolution model (section~\ref{SNR}), the CR acceleration \& diffusion model (section~\ref{acceleration}), and the shock-cloud collision model (section~\ref{shock-cloud}). 
In section~\ref{MC} and section~\ref{sea}, we justify the adopted 3D structure of nearby clouds and the density of the local CR sea, respectively.

\subsection{SNR evolution}
\label{SNR}
Summarising the multi-wavelength observations described in section~\ref{Intro}, there are three main observational features of the SNR \J: 1) a large shell structure of non-thermal X-ray and TeV emission, 2) a low density of the circumstellar medium, and 3) a hard GeV spectrum that slightly favors a leptonic $\gamma$-ray emission model. These observational features indicate that the SNR is likely expanding inside the pre-SN wind bubble with a very low density, and this could result a high shock velocity of $v_\mathrm{SNR}\gtrsim2000\mathrm{km\,s^{-1}}$\citep{Ah1999,Vink2013} at a 15\,pc radius (adopting a distance of 3.2\,kpc). 
 
The progenitor mass need to be $\gtrsim 20\mathrm{M_\odot}$ in order to blow a $>15\,$pc pre-SN wind bubble \citep{Chen2013,Chevalier1999}. There are basically two types of pre-SN bubble structures with such a massive progenitor, 1) an empty pre-SN bubble made by the main sequence (MS)/ Wolf-Rayet (WR) wind, 2) a MS bubble with a red super giant (RSG) bubble embedded inside it. A $20\mathrm{M_\odot}$ type IIL/b scenario and a $25\mathrm{M_\odot}$ type Ib/c scenario from \cite{Cui2016} are adopted in our model to represent these two bubble features.

In the $20\,\mathrm{M_\odot}$ scenario, the gas density inside the MS bubble is $n_\mathrm{MS}\sim0.01\mathrm{cm^{-3}}$ and the density inside the RSG bubble follows $n_\mathrm{RSG}(r)= \dot{M}_\mathrm{RSG}/4\pi r^2 v_\mathrm{RSG}$ \citep{Chevalier2005}, where $\dot{M}_\mathrm{RSG}$ and $v_\mathrm{RSG}$ are the mass loss rate and the wind speed during the RSG phase, respectively. By choosing  $\dot{M}_\mathrm{RSG}\approx 5\times10^{-5} \,\rm{M_{\odot}}/s$ and $v_\mathrm{RSG}\approx 15\,\rm{km/s}$ \citep{Chevalier2005}, one can obtain the RSG bubble radius as $R_\mathrm{RSG}\approx5\,$pc.

In the $25\,\mathrm{M_\odot}$ scenario, the fast H-poor WR wind is able to blow away the pre-existing RSG bubble \citep{Chevalier2005}, and leave behind a CNO core (maybe some He), which becomes the main ejecta material of a type Ib/c SN. Review work by \cite{Smartt2009} pointed out that type Ib SNe have unambiguous signatures of He and type Ic SNe show no H or He. Hence, we assume most of the H and He layers ($21\,\rm{M}_\odot$) are blown into the WR bubble with a radius of $\sim22$\,pc, we can obtain an averaged density of $n_\mathrm{WR}\sim0.02\mathrm{cm^{-3}}$ in the WR bubble. 

In both SNR scenarios, we adopt a typical CC SN explosion energy of $E_\mathrm{SN}$ as $10^{51}\mathrm{erg\,s^{-1}}$ ($\mathcal{E}_{51}$) \citep{Smartt2009}. The SN ejecta masses ($M_\mathrm{ej}$) are set as $2\,\rm{M}_\odot$ in both SNR scenarios and they are consistent with our pre-SN physics, in which the progenitor mass is the sum of the MS wind mass loss, the RSG wind mass loss, the neutron star mass ($2\,\rm{M}_\odot$), and the SN ejecta mass.

With known pre-SN environments and known SN physics, we calculate the SNR evolution history through a self-similar solution \citep{Chevalier1982,Nadezhin1985} for the ejecta-dominated stage, and a thin-shell approximation \citep{Ostriker1988,Bisnovatyi1995} for the Sedov stage. The SNR evolution parameters and results can be found in table~\ref{table:SNR}. The SNR evolution equations used here are derived by \cite{Zirakashvili2005}, they can also be found in \cite{Cui2016}.

Noticeably, dense MC-clumps, i.e.\ MC-core in our shock-cloud encountering model, can very well survive the MS wind and even the Wolf-Rayet wind, see e.g.\ the recent discovery of a dense MC-clump that survived the strong UV fields and winds from a young massive star in the 30 Doradus region \citep{Rubio2009}. 

\subsection{CR acceleration and diffusion}
\label{acceleration}
Substantial theoretical improvement has been achieved to support the idea that young SNRs can indeed accelerate particles up to 100 TeV, employing the concept of fast amplification of magnetic turbulence upstream of the SNR shocks through nonresonant streaming instability \citep{Bell2004,Zirakashvili2008}.  We adopt the analytical approximation of the MHD simulation result by \cite{Zirakashvili2008} and derive the amplified magnetic turbulence upstream and the escape energy ($E_\mathrm{max}$) by using the input parameters from the SNR evolution history. The flux of run-away CRs ($J$) and the density of the confined CRs in the acceleration region ($N_0$) are also given in \cite{Zirakashvili2008}. Here $J$ is mostly made of CRs with energies above $E_\mathrm{max}$, while $N_0$ basically follows a broken power-law with an index of -2.0 and an exponential cutoff energy of $E_\mathrm{max}$. Detailed equations about the acceleration model can be found in \cite{Zirakashvili2008,Cui2016}.

The main input parameters of our acceleration model are given through solving the SNR evolution history, they are the shock velocity -- $v_\mathrm{SNR}(t)$, the SNR radius -- $R_\mathrm{SNR}(t)$ and the density of incoming gas upstream -- $n_\mathrm{ISM}(t)$.  The other input parameters include the magnetic field in the un-shocked upstream -- $B_0$, the initial magnetic fluctuation upstream -- $B_\mathrm{b}$, and the acceleration efficiency -- $\eta_\mathrm{esc}$. Here $\eta_\mathrm{esc}$ represents the ratio between the energy flux of run-away CRs and the kinetic energy flux of incoming gas onto the shock upstream, and it is constrained by the hypothesis that the total CR energy accounts for $\lesssim10\%$ of the SN kinetic energy. A magnetic field strength of $B_0=5\mu$G in the inter-clump medium (ICM) is assumed following \cite{Crutcher2012} and a ratio of $B_\mathrm{b}/B_0=1.35\%$ is assumed following \cite{Zirakashvili2008}.

Once the CRs run away from the SNR, they enter an inhomogeneous diffusion environment, which is divided into three sub-regions: MC-core region, MC-J1729 region, and the region covering the entire outer-space other than these two MC-clumps. The last one is mostly made of the pre-SN wind bubble, the ICM and other MC structures. Inside each sub-region, a homogeneous diffusion coefficient is assumed, see also table~\ref{table:diffusion} for the adopted values of the diffusion coefficients. To calculate the diffusion process in this inhomogeneous diffusion environment, we adopt the Monte-Carlo diffusion method developed in section 2.4.2 of \cite{Cui2016}. 

\subsection{Shock-cloud collision} 
\label{shock-cloud}
When the western part of the SNR \J~encounters with the dense MC-clump, i.e.\ with MC-core, the shock is rapidly stalled \citep{Sano2010,Gabici2016} and the magnetic turbulence in the upstream and downstream of the shock is quickly damped by the high-density neutrals. This collision eventually leads to the release of all CRs confined in the shock \citep{Ohira2011}. However, \cite{Inoue2012} argued that the release of the GeV CRs may not be an immediate event after the shock-cloud collision, i.e., in case of the hadronic model of the SNR RX J1713.7-3946, \cite{Inoue2012} suggested a shell of amplified magnetic turbulence formed at the stalled shock. This turbulent shell, which could last for $\lesssim 10^3\,$year when a fast shock ($2500\,\mathrm{km\,s^{-1}}$) is hitting the MC-clump, can prevent GeV CRs from entering the MC-clumps. The relative short age of \J~(around 2 - 6\,kyr) would need to take this amplified magnetic field shell into consideration, and the effect is simplified as a delayed releasing time of the CRs ($t_\mathrm{delay} = 500$ years) after the shock-cloud collision. Furthermore, due to the short age of the SNR, the shock-cloud encounter can no longer be seen as an instantaneous event. Instead, it is described as MC-core gradually swollen by the SNR. As seen in the bottom panel of Fig~\ref{fig:scratch}, at any certain time, a belt feature will represent the shock-cloud collision area as well as the following CR leaking area.

In our numerical simulation, within a time interval $\Delta t_\mathrm{SNR}$, leaked CRs with a total number of $\Delta N_\mathrm{leak}(t) = N_0  \cdot l_\mathrm{down} \cdot A_\mathrm{belt}$ are released from the surface of the collision belt at a time of $t=t_\mathrm{SNR}+t_\mathrm{delay}$. Here $l_\mathrm{down}$ is the thickness of the acceleration region at the shock downstream which is used to normalize the total CRs trapped at the shock. $l_\mathrm{down}$ is set as $3\%R_\mathrm{SNR}$ following the MHD simulation work by \cite{Zirakashvili2008,Zirakashvili2012}, who suggested that most of the CRs and the swept gas of a young SNR are concentrated right behind the shock with a thickness $\lesssim10\%R_\mathrm{SNR}$. $A_\mathrm{belt}$ is the integrated area of the collision region during each $\Delta t_\mathrm{SNR}$, which is shown as the purple belt in Fig.~\ref{fig:scratch}. On the surface of this collision belt, the CR acceleration and CR run-away processes are immediately stopped after collision.

\subsection{Nearby molecular clouds} 
\label{MC}

\begin{figure*}[htb]
\centering
\includegraphics[height=75mm,angle=0]{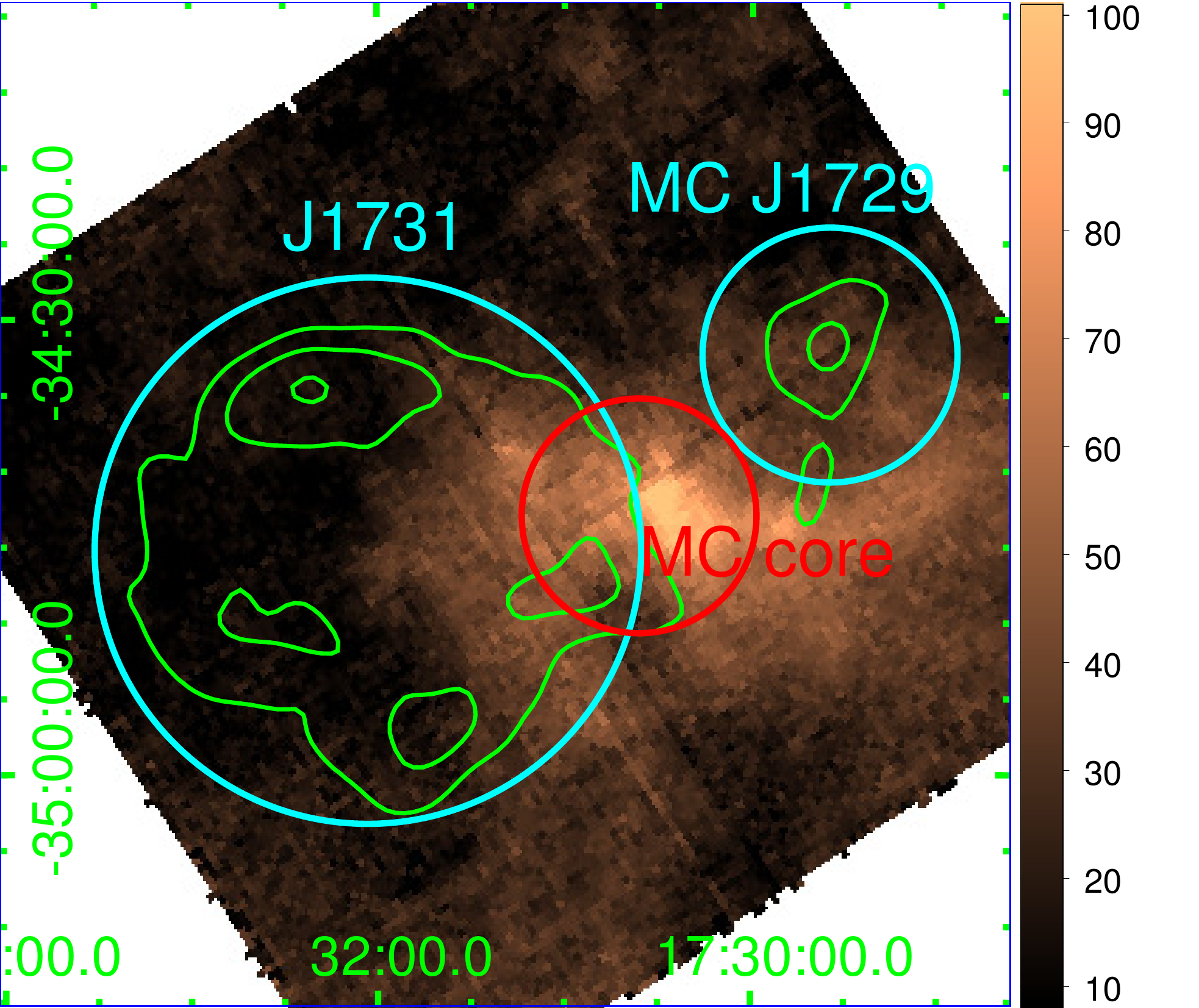}

\

\includegraphics[height=75mm,angle=0]{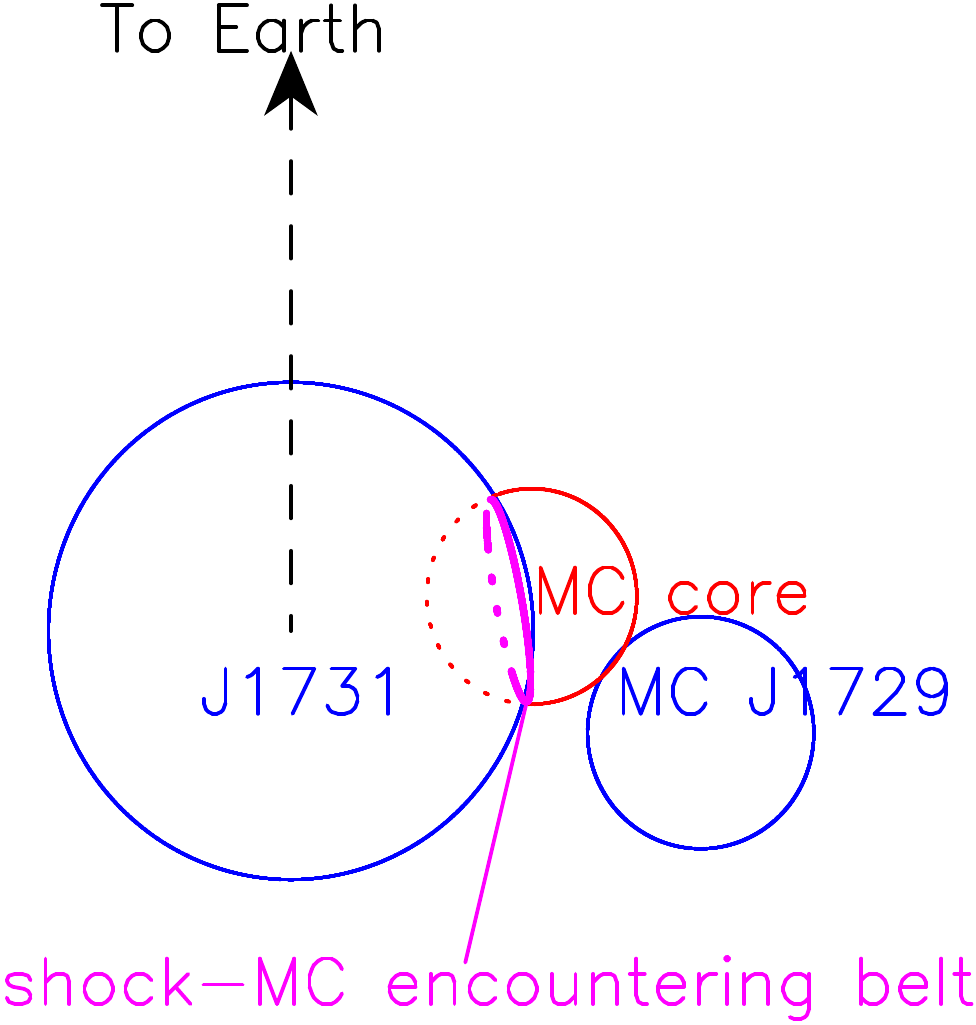}
\caption{Top panel: The CO map at \J~integrated from $-5\,\mathrm{km\,s^{-1}}$ to $-25\,\mathrm{km\,s^{-1}}$ \citep{Maxted2018}, and the TeV image by \cite{Ab2011} is shown in green contours. The color bar represents the scale of CO line intensity with a unit of $\mathrm{K\,km\,s^{-1}}$. Bottom panel: A scratch about how the SNR \J~is encountering with MC-core in our model, the 3D structures of the system -- SNR, MC-core and MC-J1729 is seen from bottom (along the Declination axis). The MC-core and MC-J1729 are spheres in our idealistic 3D structure. In both panels, the cyan and blue circles represent the SNR and MC-J1729, while the red circle represents MC-core.}
\label{fig:scratch}
\end{figure*}

From the recent $\rm{^{12}CO}$ observations with Mopra by \cite{Maxted2018}, which are shown in the top panel of Fig.~\ref{fig:scratch}, one can obtain the column density of the MC near the SNR. Our model adopts a $\rm{^{12}CO}$ emission integrated from $-5\,\rm{km/s}$ to $-25\,\rm{km/s}$ \citep[the giant MC at 3.2\,kpc,][]{Maxted2018} and a CO-to-H$_2$ mass conversion factor of $1.8\times10^{20}\,\rm{cm^{-2}K^{-1}km^{-1}s}$ \citep{Dame2001}. When compared with the previous $\rm{^{12}CO}$ observations on \J~with the CfA survey \citep{Ab2011,Cui2016}, the $\rm{^{12}CO}$ observations with Mopra have much higher angular resolution and deliver a slightly lower column density ($\sim95\%$) at MC-core region. However, at the MC-J1729 region, the CO column density with Mopra is only $\sim65\%$ of the one with CfA. We believe this is due to the low angular resolution of CfA survey, which causes contaminations into the MC-J1729 region from the bright CO features nearby. The brightest CO region with an intensity above $50\,\mathrm{K\,km\,s^{-1}}$ in the MC complex can be divided into two features, one is the MC-core, one is a bright extended CO feature at south of MC-J1729 (Western-CO-filament) which looks like a gas filament extending from MC-core to the West. These features can be seen clearer in Fig. 2 of \cite{Maxted2018}, due to a different scale of color bar used in their CO map.

Following our previous work on \J~\citep{Cui2016}, we only focus on two MC regions -- MC-core ($2.75\times 10^4\, \mathrm{M_{\odot}}$) and MC-J1729 ($1.57\times 10^4\, \mathrm{M_{\odot}}$), which are labeled as the red and blue circles in Fig.~\ref{fig:scratch}. The rest of the observed clouds in this 3.2\,kpc MC complex (excluding the MC-core and MC-J1729), especially the Western-CO-filament, are ignored in our model, due to their lack of TeV counterparts \citep{Ab2011}. The definition of MC-J1729 region is based on the on-region of HESS J1729-345 in the HESS data analysis \citep{Ab2011} (centered at $\alpha_\mathrm{J2000}$ =17h29m35s, $\delta_\mathrm{J2000} = −34^\circ$32'22'', radius $0.14^\circ$), and MC-core region represents the densest core region of this giant MC (centered at $\alpha_\mathrm{J2000}$ =17h30m36s, $\delta_\mathrm{J2000} = −34^\circ$43'0'', radius $0.13^\circ$). These two MC regions are simplified as two homogeneously filled spheres in our model. Noticeably, here the MC-core region is arbitrarily chosen following two criteria: a) The region should be large enough to cover most of the dense cloud gas around the densest core, i.e., the region with $\rm{^{12}CO}$ intensities above $50\,\mathrm{K\,km\,s^{-1}}$. b) The region should not be too large that it could overlap with HESS J1731-345 (small blue circle), because we try to separate the $\gamma$-ray emissions from these two MC clumps.

In our previous work \citep{Cui2016}, MC-core is put at a distance of 100\,pc to the SNR, hence the hadronic $\gamma$-ray emission at this region is suppressed. Recent ATLASGAL survey study \citep{Li2016} suggests that the MC-clumps inside one giant MC are more likely to be close to each other ($\lesssim$10pc) and connected through filaments. Under the assumption that the observed MC complex at 3.2\,kpc is indeed one giant MC, we use a more reasonable 3D structure in this work, as seen in the bottom panel of Fig.~\ref{fig:scratch}, MC-core and MC-J1729 are close to each other, and they are located at 3D distances of $\sim$16.91\,pc and $\sim$31.18\,pc to the SNR center, respectively. In a Cartesian frame, we set the SNR center as (x,y,z)=(0,0,0)\,pc, where (x, y) represent the distances along the directions of RA and Dec, and (z) is along the line of sight to Earth. The (x,y) values of MC-core and MC-J1729 obtained from the CO and TeV observations are (16.65, -2.14) and (28.32, -12.04), respectively. The (z) value of MC-core is chosen arbitrarily at 2 \,pc, which corresponds to a collision area of 5\% of the SNR surface at present. The (z) value of MC-J1729 is chosen as -5 \,pc, this leads to a 3D distance between MC-core and MC-J1729 of 16.83\,pc, which is roughly consistent with the ATLASGAL survey. In our model, the favored (z) value of MC-core/MC-J1729 can vary within a range of (-3,3)\,pc/(-10,10)\,pc. However, the (z) values and the diffusion coefficients ($D$) are degenerated parameters. In the model fitting, we fix the (z) value and free the $D$. Another reason that we choose these fixed (z) values is to make a relative clearer scratch figure, as shown in Fig.~\ref{fig:scratch}.

\subsection{Local CR sea} 
\label{sea}
The CR sea could play an important role in the soft GeV spectrum observed at MC-core. We adopt a radial profile of Galactic CR density \citep{Yang2016,Acero2016}, which is derived through studying the \Fermi-LAT data and the gas density in our entire galaxy. Hence, \J, with a distance to the Galactic center of around 5\,kpc, should be embedded in a CR sea with a density similar to the one observed on Earth (CR spectral index $\Gamma_\mathrm{CR,sea} \sim$ -2.55 to -2.72, energy density $U_\mathrm{CR,sea}\sim1.1\,\mathrm{eV\,cm^{-3}}$). In section~\ref{CR_sea}, we further discuss the $\gamma$-ray contributions by the CR sea.

\section{Results and discussions}
\subsection{The fixed \& fitting Parameters}
\label{parameter}
 Firstly, we summarise the fixed parameters in our model. 
\begin{itemize}
\item In our SNR model (section~\ref{SNR}), the pre-SN environment ($n_\mathrm{ISM}$) and the SNR physics ($E_\mathrm{SN}$, $M_\mathrm{ej}$) are fixed based on the multi-wavelength observations and the conventional values. An assumed value of pre-SN magnetic environment ($B_0/B_\mathrm{b}$) is chosen following \cite{Zirakashvili2008}.
\item In our acceleration and diffusion model (section~\ref{acceleration}), the escape energy ($E_\mathrm{max}$) is calculated using the nonresonant instability acceleration theory, its value is dependent on the SNR evolution history, the pre-SN environment, and the acceleration efficiency. Diffusion coefficient in outer-space ($D_\mathrm{ICM}$) is assumed as the Galactic value \citep[$D(E) = D_{10}(E/10\,\rm{GeV})^{\delta}$, $D_{10}=10^{28}\,\rm{cm^2/s},\ \delta= 0.5$,][]{Ptuskin2006}. 
\item In our shock-cloud collision model (section \ref{shock-cloud} \& \ref{MC}), the times and locations of the shock-cloud collisions are determined by the 3D MC structure and the SNR evolution history. The MC clumps are assumed to be homogeneous spheres, their 3D locations (x,y,z) are based on the CO observation and the assumed (z) values. All CRs confined in the stalled shock are assumed to be released at the collision point, and a delayed releasing time of 500\,years is assumed following the argument of \cite{Inoue2012}.
\end{itemize}

Secondly, we list the dependency of each fitting parameter below. The fitting parameters are the diffusion coefficients ($D_\mathrm{MC}$) in MC-core and MC-J1729, as well as the acceleration efficiency ($\eta_\mathrm{esc}$).  
\begin{itemize}
\item {\it The acceleration efficiency $\eta_\mathrm{esc}$}: When the acceleration efficiency increases, more kinetic energy from the incoming plasma onto the shock is transferred into CR energy. Therefore, the magnetic turbulence upstream (i.e. the escape energy) and the run-away CR flux are boosted. 
\item {\it The diffusion coefficient inside the MC-clumps $D_\mathrm{MC}$}: When the diffusion coefficient inside the MC-clumps decreases, the average time for a CR being trapped inside the MC-clumps increases, which leads to a larger accumulation of CRs at the present time. 
\end{itemize}

 In conventional models of CR diffusion around SNRs \citep[e.g.,] []{Gabici2010}, only the $D_{10}$ is set free in fitting and an one zone diffusion coefficient is adopted for the entire space. Our model uses a three zone diffusion environment (MC-core, MC-J1729, and outer-space) and we set both the  $D_{10}$ and $\delta$ free in fitting. This provides us more fitting parameters and eventually leads to an easier fitting.

\subsection{Results of the hadronic model}
\label{HadronResults}
\begin{figure*}[htb]
\centering
\includegraphics[height=10cm,angle=0]{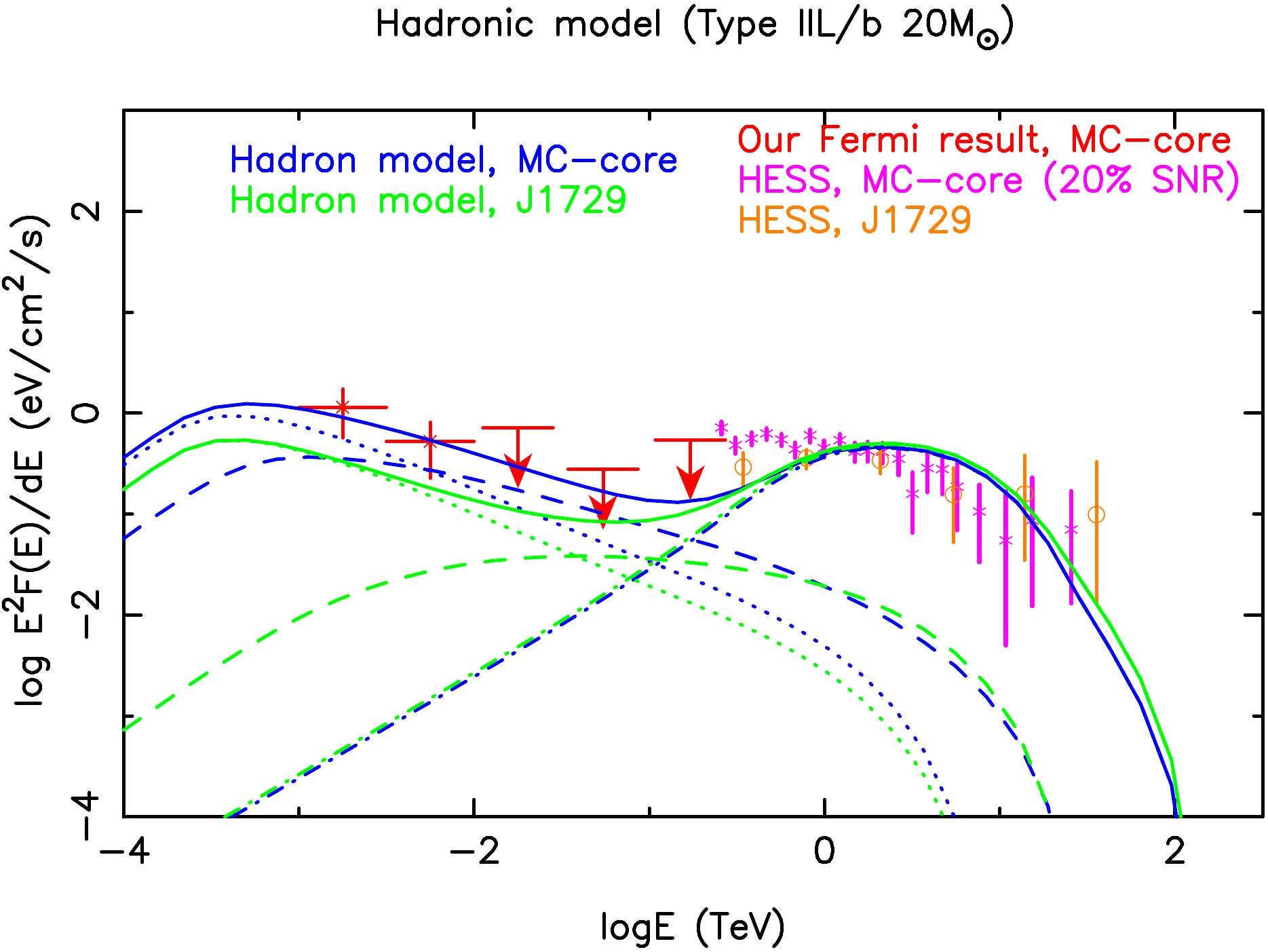} 

\

\includegraphics[height=10cm,angle=0]{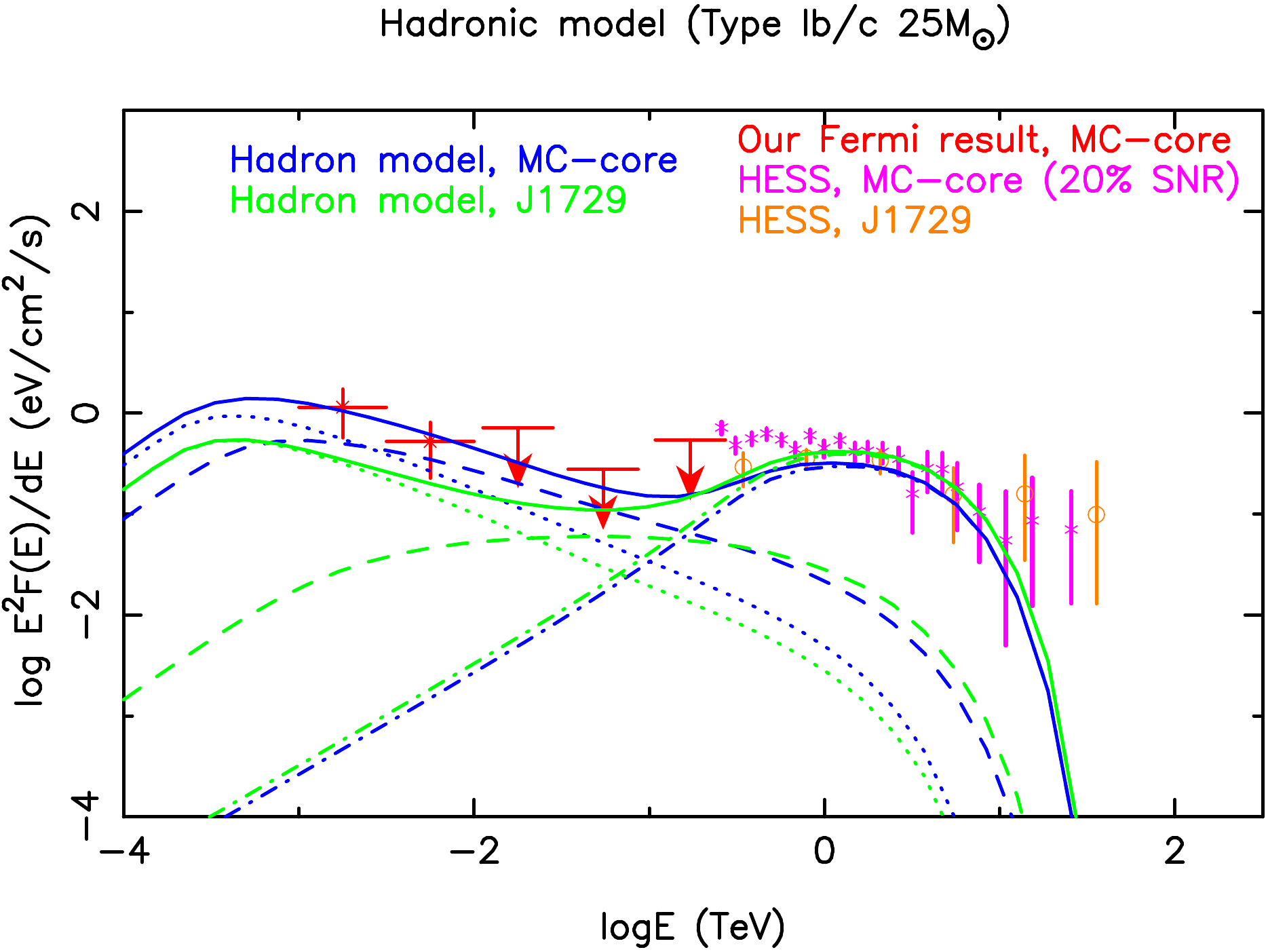} 
\caption{The hadronic model explaining the $\gamma$-ray emission at MC-core and MC-J1729. Top/bottom panels presents the model results of the SNR scenario $20\,\mathrm{M_\odot}$/$25\,\mathrm{M_\odot}$. HESS data marked in purple and orange are from \cite{Ab2011}. Our hadronic model include three components: run-away CRs from strong shocks (dashed-dotted lines), leaked CRs following the shock-cloud collision (dashed lines), and the local CR sea (dotted lines). The total of all three components are marked in solid lines.}
\label{fig:SED_hadron}
\end{figure*}

\begin{itemize}
\item {\it The SNR evolution history.} \\ In our model, the SNR with a progenitor of $20\,\mathrm{M_\odot}$/$25\,\mathrm{M_\odot}$ has spent around 6.1/2.9\,kyr expanding inside the pre-SN wind bubble before reaching 15\,pc. The western part of the SNR starts to encounter with MC-core at an age of 3.8/1.3 kyr at a radius of 9.7\,pc in the scenario of $20\,\mathrm{M_\odot}$/$25\,\mathrm{M_\odot}$. During the Sedov stage, the shock velocity is mainly dependent on the mass of the total swept gas \citep{Cui2016}. In both scenarios, $20\,\mathrm{M_\odot}$ and $25\,\mathrm{M_\odot}$, we obtain similar masses of the total swept gas at the present time ($\sim21\mathrm{M_\odot}$ in scenario $20\,\mathrm{M_\odot}$, $\sim25\,\mathrm{M_\odot}$ in scenario $25\,\mathrm{M_\odot}$), and we also obtain similar shock velocities of $v_\mathrm{SNR}\approx2100\,\mathrm{km\,s^{-1}}$ at present. In the $20\,\mathrm{M_\odot}$ scenario, after sweeping the thick RSG bubble at a radius of 5\,pc, the shock is rapidly slowed down to a velocity of $v_\mathrm{SNR}\approx2000\,\mathrm{km\,s^{-1}}$, and it maintains this high velocity during the following expansion inside the low-density MS bubble. In the $25\,\mathrm{M_\odot}$ scenario, the shock is sweeping inside the homogeneous WR bubble and its velocity is gradually reduced to $v_\mathrm{SNR}\approx2100\,\mathrm{km\,s^{-1}}$ at a radius of 15\,pc. Clearly, the averaged shock velocity in the $25\,\mathrm{M_\odot}$ scenario is much higher than the one in the $20\,\mathrm{M_\odot}$ scenario, and this leads to a younger SNR age at present. 

\item {\it The sub-regional $\gamma$-ray observations.} \\ The GeV data points of MC-core in Fig.~\ref{fig:SED_hadron} are obtained by our \Fermi-LAT analysis of the soft GeV component -- S0. No sub-regional HESS analysis of \J~has been published yet. Hence, the TeV data points of MC-core is taken as $20\%$ of the observed data points from the entire SNR region. This ratio is calculated with the TeV counts map by \cite{Ab2011}, using the TeV counts in the SNR ($F_\mathrm{SNR}$, the big blue circle in Fig.~\ref{fig:scratch}) and the counts in the MC-core region ($F_\mathrm{core}$, the red circle). Part of the SNR is overlapped by MC-core, and the TeV counts in this overlapped region account for $\sim12\%$ of $F_\mathrm{SNR}$, the rest ($88\% F_\mathrm{SNR}$) of the observed TeV emission in SNR is assumed to be leptonically dominated, see also the leptonic model in section~\ref{lepton}. Choosing a larger MC-core region will lead to a higher cloud mass and a higher sub-regional TeV flux, and vice versa.

\item {\it The spectral fitting results.} \\
In Fig.~\ref{fig:SED_hadron}, we show the $\gamma$-ray emissions from MC-core and MC-J1729 predicted in our model. In both scenarios, with progenitor masses of $20\,\mathrm{M_\odot}$ and $25\,\mathrm{M_\odot}$, the high shock velocities of $v_\mathrm{SNR}>2000\,\mathrm{km\,s^{-1}}$ during the entire SNR histories ensure that only the super-TeV CRs are able to run away from this fast shock. These run-away CRs become the main contributor to the hadronic TeV emission outside the SNR, as seen as the dash-dotted lines in Fig.~\ref{fig:SED_hadron}. The total energy of run-away CRs is $5.67\%/1.44\% E_\mathrm{SN}$ at present. When the western part of the SNR hit MC-core, the GeV CRs leaked through the stalled shock dominate the GeV emission at MC-core, as seen as the dashed lines in Fig.~\ref{fig:SED_hadron}. The local CR sea is a significant contributor to the $\lesssim10\,$GeV $\gamma$-ray emission at the MC-clumps, as seen from the dotted lines in Fig.~\ref{fig:SED_hadron}. \\
In contrast to the observed TeV spectrum from the SNR peaking at $\lesssim1$TeV, our hadronic model predicts TeV spectra from the MC-core similar to the ones from MC-J1729, which are peaking at $\gtrsim2$TeV. Future sub-regional TeV observation/analysis may resolve this MC-core region and tell us its true spectrum. The observed spectrum of MC-J1729 (index $2.24\pm0.15\mathrm{stat}$) is quite hard and shows no obvious exponential cutoff at very high energy band \citep{Ab2011}. The $25\,\mathrm{M_\odot}$ scenario fails to explain the very high energy tail ($\gtrsim20\,$TeV) of MC-J1729. The reason is that the incoming ISM density of the $25\,\mathrm{M_\odot}$ scenario remains roughly constant during the entire SNR history, in such a way the energies of run-away CRs ($E_\mathrm{max}$) are confined in a relative narrow range. In the $20\,\mathrm{M_\odot}$ scenario, run-away CR energy is boosted up to $\sim100\,$TeV during the SNR expansion inside the dense RSG bubble, which leads to a relative better fitting in the $\gtrsim20\,$TeV band. Through increasing $E_\mathrm{ej}$ or $\eta_\mathrm{esc}$ in the $25\,\mathrm{M_\odot}$ scenario, we can get a higher averaged $E_\mathrm{max}$, but this new hadronic spectrum will be still narrow and it may fail to fit the $<$1\,TeV band. The SNR evolution and CR acceleration in reality are more complex than our simplified model and they may generate a wider energy distribution of run-away CRs. 1) The SNR may encounters a clumpy WR bubble \citep{Chu1981}. 2) The acceleration efficiency could be location dependent and time dependent, see e.g., the efficient/inefficient accelerators in SNR\,1006, due to the angle of background magnetic field \citep{Petruk2009}. 

\item {\it The diffusion coefficients.} \\ In this work, we find that our best fitting results require diffusion coefficients inside the MC-clumps lower than the Galactic value, their detailed values can be found in table~\ref{table:diffusion}. Observational studies summarised by \citet{Crutcher2012} found that the maximum strength of the interstellar magnetic field stays constant at $\sim10\,\rm{\mu G}$ in the MC with densities up to $n_\mathrm{H}\sim 300\,\rm{cm}^{-3}$, and above $300\,\rm{cm}^{-3}$ it increases following a power law with exponent $\approx2/3$. Non-thermal motions (kinetic turbulence) inside dense MC-clumps are often observed as well, see e.g. the classic review by \cite{Larson1981} and the recent discovery by \cite{Li2014}. These observational findings seem to imply a low diffusion coefficient inside the dense MC-clumps, which is also consistent with previous theoretical work on CR diffusion near SNRs, in which a self-consistent picture requires low diffusion coefficients (an averaged one covering the entire space), $\sim 3-20$ times lower than typical Galactic values \citep[e.g.,] []{Gabici2007,Ah2008b,Gabici2009,Li2010,Li2012,Ohira2011,Cui2016,Cui2018}. 
\end{itemize}

In summary, we find the shock-cloud encounter scenario a plausible hypothesis to explain the $\gamma$-ray observations at MC-core and MC-J1729. Our model prefers the type IIL/b $20\,\mathrm{M_\odot}$ scenario, and diffusion coefficients lower than the Galactic one are needed inside the MC clumps.

\subsection{Leptonic $\gamma$-ray emission from the SNR other than the shock-cloud collision region}
\label{lepton}
\begin{figure*}[htb]
\centering
\includegraphics[height=10cm,angle=0]{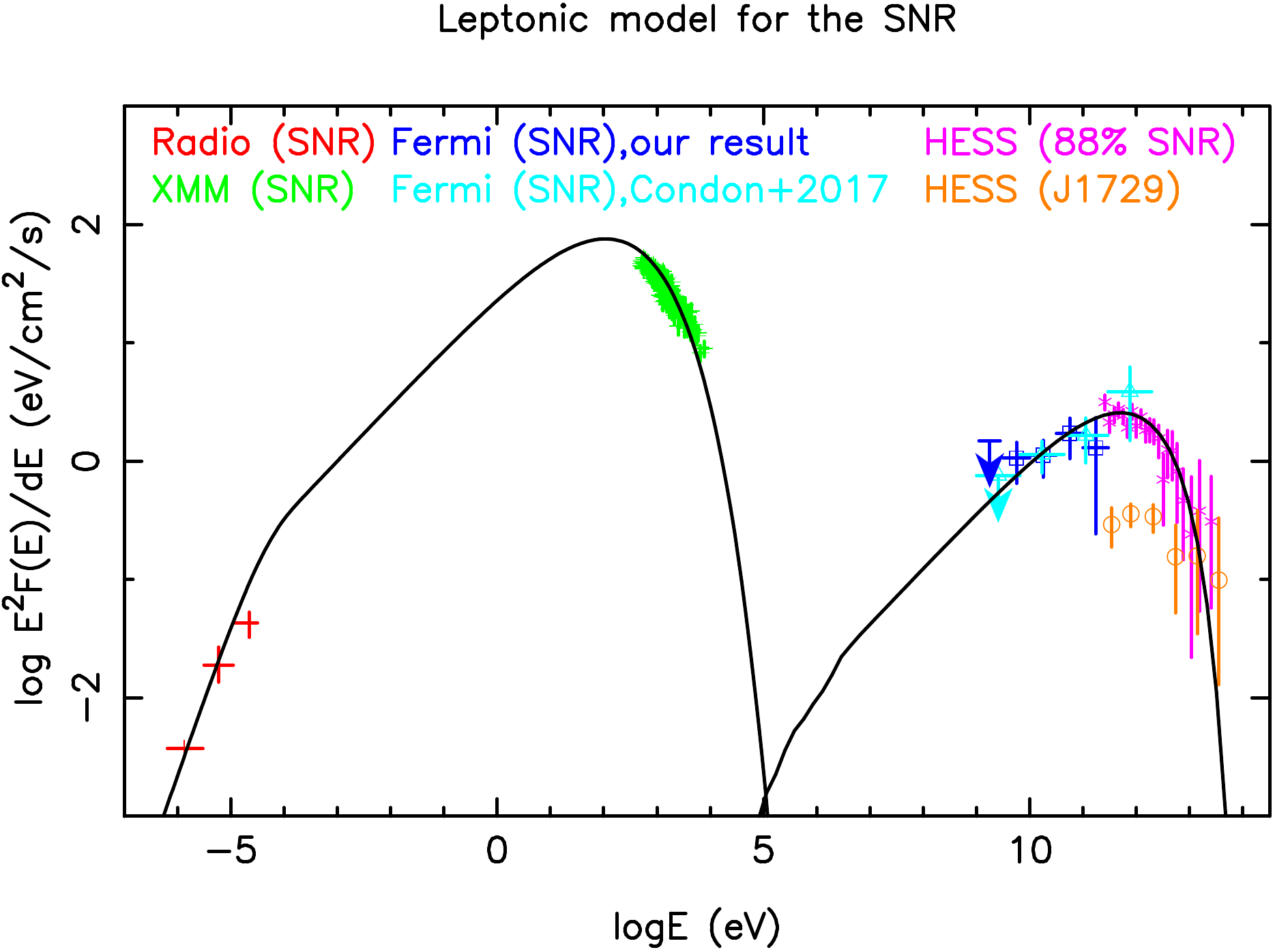} 
\caption{The leptonic model explaining the SED of the SNR (excluding MC-core region). The Radio data marked in red are from \cite{Nayana2017} (325MHz) and \cite{Tian2008} (1.4 and 5 GHz). The X-ray data represented in Green are from \cite{Doroshenko2017}. 88\% of the SNR TeV flux (excluding the MC-core region) with HESS \citep{Ab2011} is used here.}
\label{fig:SED_lepton}
\end{figure*}

Besides MC-core region, we believe that the rest of the SNR is still dominated by leptonic emission, due to the low density of the wind bubble the SNR is residing in. A hadronic attempt by \cite{Cui2016} using the swept circumstellar medium downstream as target gas has been shown to fail. In our leptonic model, as seen in Fig.~\ref{fig:SED_lepton}, the TeV data points are simply a fraction ($\sim88\%$) of the ones observed from the entire SNR (see the ``sub-regional $\gamma$-ray" paragraph in section~\ref{HadronResults} for this $\sim88\%$ fraction). The GeV data points come from our \Fermi-LAT analysis of the hard component, namely from the SNR. In our model, the fitted magnetic field downstream is $B=27\,\mathrm{\mu G}$, which is similar to the ones of RX J1713.7-3946 \citep{Abdo2011,Yuan2011}, Vela Junior \citep{Tanaka2011}, and RCW 86 \citep{Yuan2014}. We adopt a soft photon background with a temperature of 40\,K and a density of $1\,\mathrm{eV\,cm^{-3}}$, following a modified \href{http://galprop.stanford.edu}{GALPROP} model by \cite{Porter2008} for a galactocentric radius of 4 kpc. The electron population in our model follows a power-law with $\Gamma_\mathrm{e}=-2.05, E_\mathrm{cut}=10$\,TeV. A low energy cutoff at 10\,GeV is induced to better fit the radio data. The total energy of the $>1\,$GeV electrons downstream is $\sim0.042\%\mathcal{E}_{51}$, which is about 2.8\%/7.0\% of the total CR energy trapped in the shock downstream in the scenario of $20\,\mathrm{M_\odot}$/$25\,\mathrm{M_\odot}$. 

\subsection{Could the CR sea cause the soft GeV component coming from MC-core?}
\label{CR_sea}
In our hadronic model fitting, the CR sea contribution (dotted lines) at the GeV band is comparable to that of the leaked GeV CRs, especially at energies of $\lesssim10\,$GeV, see Fig.~\ref{fig:SED_hadron}. The CR sea density used in our model only represents an averaged one at a distance of $\sim5$\,kpc to the Galactic center. In reality, the density of the local CR sea, especially the GeV CRs, could vary significantly depending on the nearby environment, e.g. a star-forming region. One can also see e.g. the Galactic CR distribution simulation by \cite{Werner2015}. If the GeV emission is indeed dominated by the CR sea, we expect to see a spatial match between the \Fermi-LAT data and the CO data. 

As seen in the bottom panel of Fig.~\ref{fig:TS}, the diffuse background used in our \Fermi-LAT analysis (FL8Y) has shown a slight increase to the west of the SNR, and the flux of the diffuse background at MC-core region is around 1-2 times of the GeV flux of MC-core (S0) derived in our \Fermi-LAT analysis. If the diffuse background is overestimated, it is possible that the majority of the CR sea contribution at MC-core region has already been removed during our \Fermi-LAT background reduction and vice versa. Interestingly, our 4FGL results indeed show a lower flux at $\sim2$\,GeV.

In our work, the mass of MC-core is set to the maximum value ($2.75\times 10^4\, \mathrm{M_{\odot}}$), which is derived by integrating the CO data from $-5\,\mathrm{km\,s^{-1}}$ to $-25\,\mathrm{km\,s^{-1}}$. If we choose a smaller value of MC-core mass, the CR sea contribution can be further reduced, e.g., a CO emission integrated from $-15\,\mathrm{km\,s^{-1}}$ to $-25\,\mathrm{km\,s^{-1}}$ following \cite{Ab2011} can lead to a mass of MC-core of $1.94\times 10^4\, \mathrm{M_{\odot}}$, and a mass of MC-J1729 of $1.34\times 10^4\, \mathrm{M_{\odot}}$. To compensate the reduced hadronic $\gamma$-ray contribution from the CR sea, we could increase the contribution from leaked CRs by tuning input parameters as discussed in section~\ref{parameter}. In this work, we adopt this maximum value, which thus provides an upper limit of the CR sea contribution.

\subsection{Future observational expectations}
In this work, we separate MC-core region (S0) from the SNR in the \Fermi-LAT analysis. We expect that future spectral analysis with HESS/CTA could spatially resolve this region as well. Additionally, we also expect that future TeV observations and analysis could deliver an improved TeV image matching with the molecular clouds image at 3.2\,kpc, see e.g., the new HESS analysis attempt by \cite{Capasso2017}. We chose only two spheres -- MC-J1729 and MC-core instead of the entire MC complex into our hadronic models. Because the rest of the observed clouds in this 3.2\,kpc MC complex, especially the Western-CO-filament, lack TeV counterparts \citep{Ab2011}. Therefore, this Western-CO-filament is assumed to be located far from the SNR center ($\gg30$\,pc) in our model. However, the new HESS analysis by \cite{Capasso2017} argued that this Western-CO-filament is also shining in TeV.
 Future hard X-ray observations on this SNR could also help us to further probe the dim X-ray emission at MC-core region. 
More direct evidence for the proposed shock-cloud collision scenario would require millimeter observations, such as molecular clumps with strong velocity dispersion within the SNR, e.g., the SNR CTB\,109 \citep{Sasaki2006}, or evidence for ionisation inside the MC-clumps, e.g., in SNR W28 \citep{Vaupre2014,Maxted2016}. CC SNRs are normally considered to be born inside MCs, whereas the scenario found for \J\ with a young SNR hitting a nearby MC is clearly observationally more attractive and follow-up observations are encouraged.

\section{Conclusion}
The SNR \J, which displays a non-thermal X-ray and TeV shell structure, is believed to be still expanding inside the low-density pre-SN bubble. A dense molecular clump, called MC-core, is located at the western part of the SNR, and it is possibly embedded inside this pre-SN bubble and presently colliding with the SNR. Following the previous intriguing discoveries on \J, that MC-core region has shown a soft GeV emission (S0) and a dim X-ray emission up to 10 keV, we explored whether the SNR has collided with MC-core at its west. 
\begin{enumerate}
\item Our \Fermi-LAT analysis has unveiled two GeV components of \J, one located at the SNR center displaying a spectral index of $1.79 \pm 0.22 (stat) \pm 0.10(sys)$, and one located at MC-core displaying a spectral index of $2.42 \pm 0.22 (stat) \pm 0.10(sys)$.  We also perform a Fermi-LAT analysis using the HESS excess map as template, and its results further confirm the hard GeV component at the SNR and the soft component at MC-core.
\item We have built a hadronic model involving a shock-cloud encounter at MC-core. Our CR sources include run-away CRs from the strong shocks, leaked GeV CRs from the shock-cloud collision at MC-core region, and the local CR sea. Because of the young age of the SNR, we can not use an instantaneous event to describe the shock-cloud collision. Instead, MC-core is gradually "swallowed" by the SNR in our model. The type IIL/b $20\,\mathrm{M_\odot}$ scenario in our model can explain the GeV-TeV observations from the MC clumps, where diffusion coefficients inside the MC-clumps are about $30\%$ of the Galactic value. The multi-wavelength emissions from the rest of the SNR (other than MC-core region) are explained in a one-zone leptonic model. To better testify this hypothesis, more detailed sub-regional GeV-TeV data analysis are needed.
\item We find that leaked GeV CRs released in the shock-cloud collision are not necessarily the dominating component to explain the GeV observation at MC-core, because the CR sea with a density of $\gtrsim200\%$ of the averaged one at a galactocentric radius of 5 kpc (such density fluctuations are reasonable for the GeV CRs) can also dominate the $\lesssim10$GeV emission at MC-core.
\end{enumerate} 
 
\section*{Acknowledgement}
We like to thank Nigel Maxted and Guangxing Li for helpful discussions on the millimeter observations. We also thank the referee for the advise on the HESS template. This work is supported by the National Science Foundation of China (NSFC) grants 11633007, 11661161010, and U1731136.




\clearpage

\clearpage

\clearpage

\clearpage

\clearpage
\begin{table*}
\caption{SNR evolution of different scenarios}            
\label{table:SNR}     
\centering          
\begin{tabular}{ c c c c c c c c c c }     
\hline\hline       
SN Type \tablenote[0]{Here only $\eta_\mathrm{esc}$ is the fitting parameter, meanwhile $M$, $R_\mathrm{b,MS}$, and $R_\mathrm{b,RSG}$ are fixed.}
& $M$ \tablenote[1]{Initial mass of the progenitor star.} 
& $R_\mathrm{b,MS}$ \tablenote[2]{Size of the MS bubble (including the MS bubble shell). The numbers were chosen under the reasonable assumption that the pressure of the circumstellar medium is $10^5\,\rm{Kcm^{-3}}$ \citep{Chevalier1999,Chen2013}. The density inside the MS bubble is assumed to be as $0.01\mathrm{cm^{-3}}$ for the $20\,\rm{M_\odot}$ scenario.} 
& $R_\mathrm{b,RSG}$ \tablenote[3]{Size of RSG bubble. corresponding to $\dot{M}_\mathrm{RSG}\approx 5\times10^{-5} \,\rm{M_{\odot}}/s$ and $v_\mathrm{RSG}\approx 15\,\rm{km/s}$ in the $20\,\rm{M_\odot}$ scenario \citep{Chevalier2005}. In the $25\,\rm{M_\odot}$ scenario, the RSG is blown away by the strong WR wind and leads to a WR bubble with an averaged density of $0.02\mathrm{cm^{-3}}$.} 
& $t_\mathrm{SNR,\ end}$ \tablenote[4]{Age of the SNR when its radius expands to the size at present -- 15\,pc.} 
& $v_\mathrm{SNR,\ end}$ \tablenote[5]{Forward shock velocity of the SNR when it expands to 15\,pc.}
& $\eta_\mathrm{esc}$ \tablenote[6]{This parameter is the ratio between the energy flux from the run-away CRs and the kinetic energy flux from the upstream medium falling to the shock. } 
&$E_\mathrm{max,\ end} $ \tablenote[7]{Escape energy at the SNR shock when the shock radius reaches 15\,pc.}
&$E_\mathrm{CR, run} $ \tablenote[8]{Total energy of run-away CRs from the strong shock, integrated from the SN explosion to $t_\mathrm{SNR, \ end}$.} 
 &$E_\mathrm{CR, leak} $ \tablenote[9]{Total energy of leaked CRs from the stalled shock after the shock-cloud collision, integrated from the beginning of shock-cloud collision to $t_\mathrm{SNR, \ end}$. } \\
\hline
SNe IIL/b  &$20 \,\rm{M_{\odot}}$ &18\,pc& 5\,pc &$6.1 \,\rm{kyr}$& $2140\,\rm{km/s}$ &0.02& $34.9\, \rm{TeV}$  & $ 5.67\%\,\rm{\mathcal{E}_{51}}$ & { 0.07\%$\,\rm{\mathcal{E}_{51}}$} \\
SNe Ib/c  &$25 \,\rm{M_{\odot}}$ &22 pc& - & $2.9 \,\rm{kyr}$& $2470\,\rm{km/s}$ &0.01& $16.5\, \rm{TeV}$ & $1.44\% \,\rm{\mathcal{E}_{51}}$ & { 0.03\%$\,\rm{\mathcal{E}_{51}}$} \\
\hline                  
\end{tabular}
\end{table*}

\clearpage
\begin{table}[h!]
\caption{
Diffusion coefficients in different sub-regions ($D_{10}$ \&  $\delta$, in unit of $10^{28}\mathrm{cm^2\,s^{-1}}$ \& 1. The Galactic value is 1 \& 0.5.)}
\label{table:diffusion}
\centering
\begin{tabular}{ | c | c c c |}
\hline
SNR scenarios \tablenote[0]{The $D_{10}$ \&  $\delta$ in ICM is fixed as the galactic value, while we set the $D_{10}$ \&  $\delta$ in the two MC clumps free in fitting.} &MC-core & MC-J1729 \tablenote[1]{The MC-J1729 mass obtained with Mopra is around 65\% of that obtained in \cite{Cui2016} with CfA CO survey, hence, a lower diffusion coefficient than the one in \cite{Cui2016} is adopt here.} & ICM \tablenote[2]{Inter clump medium. In our model it represents the entire space other than the MC-clumps -- MC-core and MC-J1729.} \\ 
\hline
SNe IIL/b $20 \,\rm{M_{\odot}}$ & 0.3 \& 0.5 & 0.5 \& 0.3 & 1 \& 0.5\\
SNe Ib/c $25 \,\rm{M_{\odot}}$  & 0.9 \& 0.5 & 1.1 \& 0.3 & 1 \& 0.5\\
\hline
\end{tabular}
\end{table}

\clearpage

\clearpage

\clearpage

\end{document}